\documentclass[11pt]{article}
\usepackage{axodraw}
\usepackage{epsfig}
\usepackage{amsfonts}
\usepackage{bbm}
\input pix.sty

\newcommand{\tinymsbar}{{\overline{\mbox{\tiny\rm{MS}}}}}
\newcommand{\Lambdamsbar}{{\Lambda_\tinymsbar}}
\newcommand{\mD}{m_\rmi{D}}
\newcommand{\mG}{m_\rmi{G}}
\newcommand{\Nf}{N_{\rm f}}
\newcommand{\Nc}{N_{\rm c}}

\newcommand{\rmO}{{\mathcal{O}}}

\def\lsi{\raise0.3ex\hbox{$<$\kern-0.75em\raise-1.1ex\hbox{$\sim$}}}
\def\gsi{\raise0.3ex\hbox{$>$\kern-0.75em\raise-1.1ex\hbox{$\sim$}}}
\newcommand{\lsim}{\mathop{\lsi}}
\newcommand{\gsim}{\mathop{\gsi}}

\newcommand{\sign}{\mathop{\mbox{sign}}}

\newcommand{\nB}{n_\rmi{B}}

\newcommand{\re}{\mathop{\mbox{Re}}}

\newcommand{\Tint}[1]{{\hbox{$\sum$}\!\!\!\!\!\!\!\int\,}_{\!\!\!\!\raise-0.9ex\hbox{$\scriptstyle{#1}$}}}

\newcommand{\ZZ}{{\mathbb{Z}}}

\newcommand{\unit}{{\mathbbm{1}}} 
\newcommand{\bi}{\begin{itemize}}
\newcommand{\ei}{\end{itemize}}

\newcommand{\qq}{\tilde q_0}

\newcommand{\qqbo}{\tilde q_0} 
\def\ScatA{\picc{%
 \Laqu(20,5)(50,5)%
 \Line(10,5)(20,5)%
 \Laqu(50,4.7)(50,25.3)%
 \Lqu(10,25)(40,25)%
 \Line(40,25)(50,25)%
 \Lqu(10,4.7)(10,25.3)%
 \Lgl(30,5)(30,25)%
}}
\def\ScatB{\picc{%
 \Laqu(10,5)(50,5)%
 \Laqu(50,4.7)(50,25.3)%
 \Lqu(10,25)(50,25)%
 \Lqu(10,4.7)(10,25.3)%
 \Agl(30,25)(6,180,360)%
}}
\def\ScatC{\picc{%
 \Laqu(10,5)(50,5)%
 \Laqu(50,4.7)(50,25.3)%
 \Lqu(10,25)(50,25)%
 \Lqu(10,4.7)(10,25.3)%
 \Agl(30,5)(6,0,180)%
}}
\def\ScatD{\picc{%
 \Laqu(10,5)(50,5)%
 \Laqu(50,4.7)(50,25.3)%
 \Lqu(10,25)(50,25)%
 \Lqu(10,4.7)(10,17)%
 \Line(10,17)(10,25.3)%
 \Agl(10,25)(8,270,360)%
}}
\def\ScatE{\picc{%
 \Laqu(10,5)(50,5)%
 \Laqu(50,4.7)(50,25.3)%
 \Lqu(10,25)(50,25)%
 \Lqu(10,15)(10,25.3)%
 \Line(10,4.7)(10,15)%
 \Agl(10,5)(8,0,90)%
}}
\def\ScatF{\picc{%
 \Laqu(10,5)(50,5)%
 \Line(50,4.7)(50,15)%
 \Laqu(50,15)(50,25.3)%
 \Lqu(10,25)(50,25)%
 \Line(10,15)(10,25.3)%
 \Lqu(10,4.7)(10,15)%
 \Lgl(10,15)(50,15)%
}}
\def\ScatG{\picc{%
 \Laqu(10,5)(50,5)%
 \Laqu(50,4.7)(50,25.3)%
 \Lqu(10,25)(50,25)%
 \Lqu(10,4.7)(10,25.3)%
 \Agl(10,15)(6,270,90)%
}}
\def\ScatH{\picc{%
 \Laqu(10,5)(50,5)%
 \Laqu(50,4.7)(50,25.3)%
 \Lqu(10,25)(50,25)%
 \Lqu(10,4.7)(10,25.3)%
 \Agl(50,15)(6,90,270)%
}}
\def\ScatI{\picc{%
 \Laqu(10,5)(50,5)%
 \Laqu(50,4.7)(50,17)%
 \Lqu(10,25)(50,25)%
 \Lqu(10,4.7)(10,25.3)%
 \Line(50,17)(50,25.3)%
 \Agl(50,25)(8,180,270)%
}}
\def\ScatJ{\picc{%
 \Laqu(10,5)(50,5)%
 \Laqu(50,13)(50,25.3)%
 \Lqu(10,25)(50,25)%
 \Lqu(10,4.7)(10,25.3)%
 \Line(50,4.7)(50,13)%
 \Agl(50,5)(8,90,180)%
}}
\newcommand{\picmin}[1]{\PIC{#1}{\pwcc}{\phgt}{1.0}}
\def\ScatFmin{\picmin{%
 \Laqu(10,5)(50,5)%
 \Line(50,4.7)(50,15)%
 \Laqu(50,15)(50,25.3)%
 \Lqu(10,25)(50,25)%
 \Line(10,15)(10,25.3)%
 \Lqu(10,4.7)(10,15)%
 \Lgl(10,15)(50,15)%
}}

\makeatletter \@addtoreset{equation}{section} \makeatother
\renewcommand{\theequation}{\arabic{section}.\arabic{equation}}
\makeatletter
\renewcommand\section{\@startsection {section}{1}{\z@}%
                                   {-5.5ex \@plus -1ex \@minus -.2ex}
                                   {2.3ex \@plus.2ex}%
                                   {\normalfont\large\bfseries}}
\renewcommand\subsection{\@startsection{subsection}{2}{\z@}%
                                     {-3.25ex\@plus -1ex \@minus -.2ex}%
                                     {1.5ex \@plus .2ex}%
                                     {\normalfont\normalsize\bfseries}}
\renewcommand\thesection {\@arabic\c@section}
\renewcommand\thesubsection   {\thesection.\@arabic\c@subsection}
\renewcommand{\@seccntformat}[1]{%
\csname the#1\endcsname.\hspace{1.0em}}
\makeatother

\begin{document}

\begin{titlepage}
\begin{flushright}
BI-TP 2006/41\\
MS-TP-06-32\\
INT PUB 06-37\\
hep-ph/0611300\\ \vspace*{1cm}
\end{flushright}
\begin{centering}
\vfill

{\Large{\bf Real-time static potential in hot QCD}} 

\vspace{0.8cm}

M.~Laine$^\rmi{a}$, 
O.~Philipsen$^\rmi{b}$, 
P.~Romatschke$^\rmi{c}$, 
M.~Tassler$^\rmi{b}$ 

\vspace{0.8cm}

$^\rmi{a}${\em
Faculty of Physics, University of Bielefeld, 
D-33501 Bielefeld, Germany\\}

\vspace*{0.3cm}
 
$^\rmi{b}${\em
Institute for Theoretical Physics, University of M\"unster, 
D-48149 M\"unster, Germany\\}

\vspace*{0.3cm}

$^\rmi{c}${\em
INT, University of Washington, 
Box 351550, Seattle WA, 98195, USA\\} 

\vspace*{0.8cm}

\mbox{\bf Abstract}
 
\end{centering}

\vspace*{0.3cm}
 
\noindent
We derive a static potential for a heavy quark-antiquark pair 
propagating in Minkowski time at finite temperature, by defining
a suitable gauge-invariant Green's function and computing it
to first non-trivial order in Hard Thermal Loop resummed perturbation 
theory. The resulting Debye-screened potential could be used in 
models that attempt to describe the ``melting'' of heavy quarkonium
at high temperatures. We show, in particular, that the potential develops 
an imaginary part, implying that thermal effects generate a finite width 
for the quarkonium peak in the dilepton production rate. 
For quarkonium with a very heavy constituent mass $M$, the width
can be ignored for $T \lsim g^2 M/12\pi$, where $g^2$ is the strong gauge 
coupling; for a physical case like bottomonium, it could become
important at temperatures as low as 250 MeV. Finally, we point out that 
the physics related to the finite width originates from the Landau-damping
of low-frequency gauge fields, and could be studied non-perturbatively 
by making use of the classical approximation.

\vfill

 
\vspace*{1cm}
  
\noindent
February 2007

\vfill

\end{titlepage}

%
\section{Introduction}

Heavy quarkonium systems, $c\bar c$ and $b\bar b$, have turned out 
to provide extremely useful probes for QCD phenomenology~\cite{quarkonium}. 
On the theoretical side, the existence of a heavy mass scale makes these
systems more susceptible to analytic treatments than hadrons
made out of light quarks only, while on the experimental side, 
decays of heavy quarkonia lead to relatively clean signals which 
are precisely measured by now. It is therefore not surprising that
the modifications that the quarkonium systems may undergo at high
temperatures, are also among the most classic observables 
considered for heavy ion collision experiments~\cite{ms}.

Let us recall explicitly the conceptually clean connection 
that heavy quarkonium provides between thermal field theory
and heavy ion phenomenology. Consider the 
production rate $\Gamma_{\mu^+\mu^-}$ 
of dileptons (for example, $\mu^+\mu^-$ pairs) 
with a total four-momentum $Q = P_{\mu^+} + P_{\mu^-}$ from 
a hot thermal medium. It can be shown that this rate 
is given by~\cite{dilepton}
\be
 \frac{{\rm d} \Gamma_{\mu^+\mu^-}}{{\rm d}^4 Q} = 
 -\frac{e^2}{3 (2\pi)^5 Q^2} 
 \biggl( 1 + \frac{2 m_\mu^2}{Q^2}
 \biggr)
 \biggl(
 1 - \frac{4 m_\mu^2}{Q^2} 
 \biggr)^\fr12 \eta_{\mu\nu} \tilde C_{<}^{\mu\nu}(Q)
 \;, 
 \la{dilepton}
\ee
where we assumed $Q^2 \ge (2 m_\mu)^2$, 
$e$ is the electromagnetic coupling, 
$\eta_{\mu\nu} = \mathop{\mbox{diag}}$($+$$-$$-$$-$), 
and $\tilde C_{<}$ is 
a certain two-point correlator of the electromagnetic current 
$\hat \mathcal{J}^\mu(x)$ in the Heisenberg picture, 
$\hat \mathcal{J}^\mu(x) = ... 
 + \fr23 \! e \; \hat{\!\bar c}\, (x) \gamma^\mu \hat{c}(x) 
 - \fr13 \! e \, \hat{\bar b}\, (x) \gamma^\mu \hat{b}(x)$: 
\be
 \tilde C_{<}^{\mu\nu}(Q) 
 \equiv 
 \int_{-\infty}^\infty 
 \! {\rm d}t 
 \int \! {\rm d}^3 \vec{x}\,
 e^{i Q\cdot x}
 \Bigl\langle
  \hat \mathcal{J}^\nu (0) 
  \hat \mathcal{J}^\mu (x)
 \Bigr\rangle
 \;. \la{smaller}
\ee
The expectation value refers to 
$\langle...\rangle\equiv \mathcal{Z}^{-1} \tr [\exp(-\hat H/T)(...)]$, 
where $\mathcal{Z}$ is the partition function, 
$\hat H$ is the QCD Hamiltonian operator, and $T$ is the temperature.

Now, the heavy quark parts in the electromagnetic current
induce a certain structure into the dilepton production rate. 
More precisely, they produce a {\em threshold} 
at around $Q^2 \simeq (2 M)^2$, where $M$ is the 
heavy quark mass, and a {\em resonance peak} (or peaks) near 
the threshold, with a certain height and width. The height and 
width could in principle be observed, as a function of 
the temperature $T$ that is reached in the collision. 
Of course there are all kinds of practical limitations to this, 
related to non-equilibrium features, 
background effects, the energy resolution of the detector, etc,  
but at least some broad features like a total disappearance
(``melting'') of the quarkonium peak should ultimately be visible. 

These circumstances have lead to a great number 
of studies of quarkonium physics at high temperatures. 
For instance, various types of potential models have been 
developed~\cite{sd}, with non-perturbative input taken from lattice
simulations~\cite{kz}. 
It is not quite clear, however, to which extent potential 
models are appropriate at finite temperatures, or which
non-perturbative potentials should be used as input:
at asymptotically large distances, 
the standard Polyakov loop correlator is known to fail 
to reproduce the expected Debye-screened potential~\cite{MDNLO},
while many modified descriptions are
afflicted by gauge ambiguities~\cite{jp}.
More recently, 
direct lattice determinations of the quarkonium spectral function
have been attempted~\cite{th}, 
given that in principle Euclidean data {\em can} be 
analytically continued to Minkowski spacetime~\cite{acont}
(though in practice model assumptions need again to be introduced, 
given the very finite number of points in time
direction and the statistical nature of the data that are 
available for lattice studies). Finally, similar observables
have been considered for strongly-coupled $\mathcal{N} = 4$
Super-Yang-Mills theory, through the AdS/CFT correspondence~\cite{ads}.

All of the studies mentioned, however, either resort to some
degree of modelling, or attempt to tackle the complete
problem by ``brute force'' on the lattice. It is our philosophy here
to rather make some more use of the existence of the heavy mass $M$
and the high temperature $T$ that characterise the system, and the 
property of asymptotic freedom of QCD. Indeed, the strict weak-coupling
expansion does appear to be applicable (within 
10--20\% accuracy) to hot QCD at temperatures
as low as a few hundred MeV, once worked out to a high enough order and
supplemented possibly by numerically determined non-perturbative
{\em coefficients}, rather than complete functions; 
evidence for this has been obtained through precision studies
with a large number of independent observables, such as spatial 
correlation lengths~\cite{ah,lv}, 
the spatial string tension~\cite{gE2,lattsigmas}, 
quark number susceptibilites~\cite{bir,av,lattsusc}, 
the 't Hooft loop tension~\cite{pg,pdf}, 
and perhaps also the equation-of-state~\cite{pheneos}.

More precisely, our goal here is to derive, through a resummed
perturbative computation, a static potential that a heavy 
quark-antiquark pair propagating in Minkowski time at a finite
temperature feels. This relatively simple computation illustrates a few 
interesting phenomena that have not been exhaustively addressed before, 
as far as we know. Eventually, our goal is to take the lessons
brought by this study to a practical level, by carrying out a more 
extensive numerical investigation 
of the properties of heavy quarkonium~\cite{prg}.

The outline of this note is as follows. 
In \se\ref{se:basic} we set up the observable to be 
determined. The actual computation is discussed 
in some detail in \se\ref{se:details}. We elaborate
on the main results in \se\ref{se:discussion}, and conclude 
in \se\ref{se:concl}. There are two appendices summarising 
well-known formulae from the literature with our conventions. 

%
\section{Basic setting}
\la{se:basic}

Rather than \eq\nr{smaller}, we prefer to concentrate in the following
on the correlator
\be
 \tilde C_{>}(Q) 
 \equiv \eta_{\mu\nu}\,
 \int_{-\infty}^\infty 
 \! {\rm d}t 
 \int \! {\rm d}^3 \vec{x}\,
 e^{i Q\cdot x}
 \Bigl\langle
  \hat \mathcal{J}^\mu (x)
  \hat \mathcal{J}^\nu (0) 
 \Bigr\rangle
 \;. \la{larger}
\ee
As we recall in Appendix~A, 
the two time orderings are related to each other by 
$\tilde C_{<}(Q) = \exp(-\beta q^0) \tilde C_{>}(Q)$; 
for $q^0\gsim 2M$, $\tilde C_{>}(Q)$ is of order unity
while $\tilde C_{<}(Q)$ is exponentially suppressed.
Furthermore, let us for simplicity set the spatial momentum to zero, 
$\vec{q} = \vec{0}$, and consider the correlator before taking the 
Fourier transform with respect to time:
\be
 C_{>}(t) 
 \equiv 
 \int \! {\rm d}^3 \vec{x}\,
 \Bigl\langle
  \hat \mathcal{J}^\mu (t,\vec{x})
  \hat \mathcal{J}_\mu (0,\vec{0}) 
 \Bigr\rangle
 \;. \la{simplelarger}
\ee
Note that translational invariance 
guarantees that $C_{>} (-t) = C_{<} (t)$.

Now, the determination of $\tilde C_{>}(Q)$ around the threshold
requires a resummation of the perturbative series (this is the case
even at zero temperature). A way to implement this is to deform 
the correlator suitably so that we obtain a Schr\"odinger-type equation, 
which can then be solved ``non-perturbatively'', implementing 
an all-orders resummation. To arrive at a Schr\"odinger equation, 
we define a point-splitting, by introducing
a vector $\vec{r}$, and consider an extended interpolating operator
for the electromagnetic current, rather than a local one
(we also drop the electromagnetic couplings, 
and denote $\hat{c},\hat{b}$
by a generic quark field $\hat\psi$): 
\be
 \check C_{>}(t,\vec{r}) 
 \equiv 
 \int \! {\rm d}^3 \vec{x}\,
 \Bigl\langle
  \hat{\!\bar\psi}\,(t,\vec{x}+\frac{\vec{r}}{2})
  \gamma^\mu
  \, W_{\vec{r}}[(t,\vec{x}+\frac{\vec{r}}{2});
                         (t,\vec{x}-\frac{\vec{r}}{2})] \, 
  \hat \psi(t,\vec{x}-\frac{\vec{r}}{2}) \;\; 
  \hat{\!\bar\psi}\,(0,\vec{0})
  \gamma_\mu
  \hat{\psi}(0,\vec{0})
 \Bigr\rangle
 \;. \la{poinsplit}
\ee
Here $W_{\vec{r}}[x_1,x_0]$ is a Wilson line from $x_0$ 
to $x_1$, along a straight path in the direction of $\vec{r}$, 
inserted in order to keep the interpolating operator gauge-invariant.
Once the solution for $\check C_{>}(t,\vec{r})$ is known, 
one can return back to the original situation $\vec{r} = \vec{0}$.
We stress that only the $\vec{r} = \vec{0}$ limit corresponds 
to the physical current-current correlator in \eq\nr{simplelarger}
that we are interested in.  
Indeed, the way the point-splitting is carried out may not be unique, 
but these ambiguities should disappear once we set 
$\vec{r} = \vec{0}$ in the solution. 

Let us inspect $\check C_{>}(t,\vec{r})$ in the limit 
that the heavy quark mass, $M$, becomes very large. Then the 
heavy quarks are essentially non-relativistic, and should 
also be weakly interacting. Ignoring interactions altogether,
a straightforward diagrammatic computation shows that  
the correlator in this limit indeed satisfies a Schr\"odinger-type equation, 
\be
 \biggl[ i \partial_t - \biggl( 2 M - \frac{\nabla_\vec{r}^2}{M}+ 
 \rmO\Bigl( \frac{1}{M^3} \Bigr)
 \biggr) \biggr]  \check C_{>}(t,\vec{r}) = 0 
 \;,  \la{Seq}
\ee
with the initial condition 
$
 \check C_{>}(0,\vec{r}) = - 6 \Nc\, \delta^{(3)}(\vec{r})
$.
Here the factor 6 corresponds physically to two heavy spin-1/2 degrees 
of freedom times a sum over the spatial components of the current.
For $\check C_{<}(t,\vec{r})$, 
an analogous computation in the same limit yields
\be
 \biggl[ i \partial_t + \biggl( 2 M - \frac{\nabla_\vec{r}^2}{M}+
 \rmO\Bigl( \frac{1}{M^3} \Bigr)
 \biggr) \biggr]  \check C_{<}(t,\vec{r}) = 0 
 \;,  \la{Seqprime}
\ee
with the same initial condition; this form is consistent with 
the symmetry $C_{>} (-t) = C_{<} (t)$.

Once interactions are switched on, we might expect (apart from 
the renormalization of the parameter $M$ and of the initial condition)
to generate a potential, $V(\vec{r})$, into \eqs\nr{Seq}, \nr{Seqprime}, 
appearing in a form familiar from non-relativistic quantum mechanics.
To be precise, we define the potential to be a function inside the round 
parentheses in \eq\nr{Seq} which scales as $M^0$ in the large-$M$ expansion.  
To simplify the situation a bit, we would like to be able to  
extract the potential directly, 
without needing to bother about gradients like $\nabla_{\vec{r}}^2/M$. 
One possibility would be to consider the modified object
\ba
 \bar C_{>}(t,\vec{r}) 
 & \equiv & 
 \Bigl\langle
  \hat{\!\bar\psi}\,(t,\frac{\vec{r}}{2})
  \gamma^\mu
  \, W_{\vec{r}}[(t,\frac{\vec{r}}{2});
                         (t,-\frac{\vec{r}}{2})] \, 
  \hat \psi(t,-\frac{\vec{r}}{2}) 
  \times \nn & & \hphantom{\frac{4}{9} e^2 \,
  \Bigl\langle} \times 
  \hat{\!\bar\psi}\,(0,-\frac{\vec{r}}{2})
  \gamma_\mu
  \, W_{-\vec{r}}[(0,-\frac{\vec{r}}{2});
                         (0,\frac{\vec{r}}{2})] \, 
  \hat{\psi}(0,\frac{\vec{r}}{2})
 \Bigr\rangle
 \;. \la{barC}
\ea
Here the quarks can be considered infinitely heavy, such that 
they do not move in space; the integral over $\vec{x}$ has consequently 
been dropped; and the quark fields at $t=0$ point-split. 

To simplify the situation even further, one notes that 
the infinitely heavy quark propagators are themselves represented
by Wilson lines, apart from the ``trivial'' phase factor $\exp(-iMt)$. 
Thereby we arrive at a Wilson loop. It should be noted, though, that 
time ordering still plays a role, and the precise specification
of the object considered is not as simple as in Euclidean field theory. 
In fact, the Minkowskian Wilson loop is most naturally defined just as 
an appropriate analytic continuation of the Euclidean object. 

To summarise, we can imagine at least two ways of defining what we might
call the static potential in real time. 
On one hand, one can start by computing the 
standard Wilson loop in Euclidean spacetime, and then carry out the 
analytic continuation that leads to the time ordering in \eq\nr{barC}.
This is the computation that will be described below.
On the other hand, one can consider directly \eq\nr{barC}, 
just replacing the quarks by (almost) infinitely heavy static ones:
this amounts to NRQCD~\cite{nrqcd} with the omission of terms
of order $\rmO(1/M)$. In the heavy-quark limit,
$\bar C_{>}(t,\vec{r})$ is dominated by the 
forward-propagating part, with the phase 
factor $\sim \exp(- 2 i Mt)$; the static potential 
can then be extracted from an equation like \eq\nr{Seq}. 
We have carried out this computation as well, and  
do indeed obtain the same result as from the analytic continuation 
of the Euclidean Wilson loop; however, given that the NRQCD computation 
is quite a bit more involved than the Wilson loop one, 
we concentrate on the latter in the following.
 
As indicated above, the calculation of a static potential necessarily
involves a point splitting and thus should be regarded as an intermediate
step towards the computation of a current-current correlator. 
This feature is obviously shared by potential model approaches. 
However, we believe that our procedure offers several advantages over
the latter. Since our calculation is performed in a well-defined field
theoretic setting, the connection with the physical real-time observable 
of interest remains manifest, while non-perturbative gauge invariance as well 
as the correct perturbative limit of our static potential are ensured. 

%
\section{Details of the computation}
\la{se:details}

Since we make use both of Minkowskian and Euclidean
metrics, let us start by introducing some notation to keep
them apart. Minkowskian four-momenta are denoted by capital
symbols, $Q$, with components $q^\mu$, while Euclidean ones
are denoted by $\tilde Q$, with components $\tilde q_\mu$. 
The indices are kept down in the latter case, and our convention is  
$\tilde q_\mu \equiv(\qqbo,\tilde q_i) \equiv (\qqbo,-q^i)$. 
Spacetime coordinates are denoted by $x,\tilde x$, and
here our convention is $\tilde x_\mu \equiv (\tilde x_0, x^i)$.
The Euclidean scalar product is thereby naturally defined as 
$\tilde x\cdot \tilde Q \equiv \tilde x_\mu\tilde q_\mu =
\tilde x_0 \tilde q_0 - x^i q^i$,  
the four-volume integral as 
$\int_{\tilde x} \equiv \int_0^\beta \! {\rm d} \tilde x_0  
 \int \! {\rm d}^3\vec{x}$, and the thermal sum-integral
as $\,\int\!\!\!\!\!{\raise0.1em\hbox{$\scriptstyle\sum$}}_{\tilde Q} = 
T \sum_{\qqbo} \int \! {\rm d}^3 \vec{q}/(2\pi)^3$.
Wick rotation amounts to $\tilde x_0 \leftrightarrow i x^0$, 
$\qq \leftrightarrow - i q^0$.
All Matsubara frequencies we will meet are bosonic: 
i.e. $\qqbo = 2 \pi n T$, $n \in \ZZ$. The temperature
is often expressed as $\beta \equiv T^{-1}$. 

%
\begin{figure}[t]
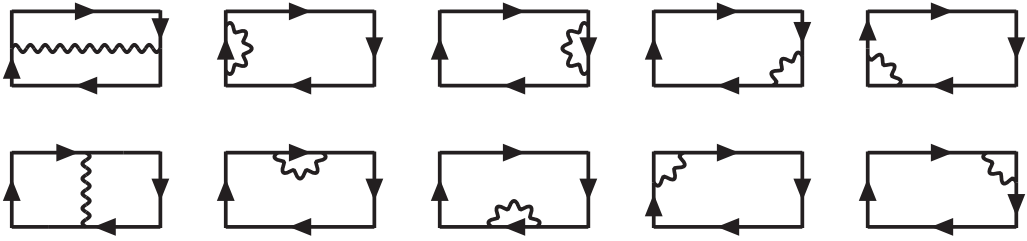


\begin{eqnarray*}
&& 
 \hspace*{-2cm}
 \ScatF \qquad\quad 
 \ScatG \qquad\quad 
 \ScatH \qquad\quad 
 \ScatJ \qquad\quad 
 \ScatE \\[10mm] 
&& 
 \hspace*{-2cm}
 \ScatA \qquad\quad 
 \ScatB \qquad\quad 
 \ScatC \qquad\quad 
 \ScatD \qquad\quad 
 \ScatI
\end{eqnarray*}

\caption[a]{\small 
The graphs contributing to the static potential at $\rmO(g^2)$. Arrows
indicate heavy quarks or Wilson lines, and wiggly lines stand for gluons.} 
\la{fig:graphs}
\end{figure}
%

\subsection{Wilson loop with Euclidean time direction}

Let again $W[\tilde z_1;\tilde z_0]$ be a Wilson line from 
point $\tilde z_0$ to point $\tilde z_1$: 
\be 
 W[\tilde z_1;\tilde z_0] = \unit + 
   ig 
   \int_{\tilde z_0}^{\tilde z_1} \! {\rm d} \tilde x_\mu A_\mu(\tilde x)
 + (ig 
       )^2 
          \int_{\tilde z_0}^{\tilde z_1} \! {\rm d} \tilde x_\mu  
          \int_{\tilde z_0}^{\tilde x} \! {\rm d} \tilde y_\nu  \,
          A_\mu(\tilde x) A_\nu(\tilde y) 
 + \ldots
 \;,  
\ee
where $A_\mu = A_\mu^a T^a$ and $T^a$ are the Hermitean generators
of SU($\Nc$), normalised as $\tr[T^a T^b] = \delta^{ab}/2$.
The Euclidean correlation function considered is then defined as
\be
 C_E(\tau,\vec{r}) \equiv 
 \frac{1}{\Nc} \tr 
 \Bigl\langle
  W[(0,\vec{r});(\tau,\vec{r})]\;
  W[(\tau,\vec{r});(\tau,\vec{0})]\;
  W[(\tau,\vec{0});(0,\vec{0})]\;
  W[(0,\vec{0});(0,\vec{r})]                                            
 \Bigr\rangle
 \;, 
\ee
where we have for convenience shifted the origin 
by $\vec{r}/2$ with respect to \eq\nr{barC}. The prefactor $1/\Nc$
has been inserted as a normalization,  
guaranteeing that $C_E(0,\vec{r}) = 1$.

We can formally expand $C_E$ in a power series in the coupling 
constant $g^2 
$, understanding of course that the infrared problems
of finite-temperature field theory necessitate the use of resummed
propagators in order for this procedure to be valid (cf.\ Appendix B): 
$C_E = C_E^{(0)} + C_E^{(2)} + ...\;$, where the 
superscript indicates the power of $g$ appearing as a prefactor. 
The leading order result is trivial, $C_E^{(0)} = 1$. 
We now turn to the computation of $C_E^{(2)}$.
The graphs entering at this order
are shown in~\fig\ref{fig:graphs}.

The computation of the graphs in~\fig\ref{fig:graphs}
is not quite trivial. The problem is 
that in order to avoid ambiguous expressions (of the type ``$0/0$''), 
one needs to treat the Matsubara zero-modes very carefully. In fact, 
we treat them separately from the non-zero modes. 
For the $\rmO(g^2)$-contribution to the 
first graph in \fig\ref{fig:graphs}, 
for instance, we obtain
\ba
 \ScatFmin\quad & = &  g^2 C_F 
 \int_0^\tau \! {\rm d} \tilde x_0
 \int_0^\tau \! {\rm d} \tilde y_0 \,
 \Tint{\tilde Q} e^{i\tilde q_0 (\tilde x_0 - \tilde y_0) - i q^3 r}
 \biggl[
   \frac{P^E_{00}(\tilde Q)}{\tilde Q^2 + \Pi_E(\tilde Q)} + 
   \xi \frac{(\tilde q_0)^2}{(\tilde Q^2)^2} 
 \biggr]
 \nn   
 & = & 
 g^2 C_F \int \! \frac{{\rm d}^3\vec{q}}{(2\pi)^3}
 \Bigl( e^{-iq^3 r} \Bigr)
 \biggl\{ 
 \tau^2 T \; \frac{P^E_{00}(0,\vec{q})}{\vec{q}^2 + \Pi_E(0,\vec{q})}
 + 
 \nn & + & 
 T \sum_{\qqbo\neq 0}
 \frac{
   2 - e^{i \qqbo\tau} - e^{- i \qqbo \tau} 
 }{\qqbo^2}
 \biggl[ 
 \frac{P^E_{00}(\tilde Q)}{\tilde Q^2 + \Pi_E(\tilde Q) }
   + \xi \frac{(\tilde q_0)^2}{(\tilde Q^2)^2} 
 \biggr] 
 \biggr\}
 \;, \la{graph1}
\ea
where we have inserted the gluon propagator from \eq\nr{prop}; 
defined $C_F \equiv (\Nc^2 -1)/2\Nc$;  
and chosen the vector $\vec{r}$
to point in the $x^3$-direction: $\vec{r}\equiv (0,0,r)$.
It is perhaps appropriate to note that if one would rather 
attempt to keep all the Matsubara modes together, as an expression
of the type on the latter row in \eq\nr{graph1}, and then try to carry 
out the sum with the usual contour trick~\cite{leb,kg}, the zero-mode
$\tilde q_0 = 0$ is not treated correctly, because of the ambiguity 
of the expression under $\tau\to \tau+ n\beta$, $n\in\ZZ$.

Summing all the graphs together, the 
parts proportional to $\xi$, as well as the 
``longitudinal parts'' $\propto \tilde q_\mu\tilde q_\nu$ from 
the terms multiplied by $P^E_{\mu\nu}$
(cf.\ \eq\nr{PE}), cancel explicitly. 
Inserting finally the expressions of the projection operators
from \eqs\nr{PT}, \nr{PE}, we obtain
\ba
 C_E^{(2)}(\tau,r) 
 \!\! & = & \!\! g^2 C_F \int \! \frac{{\rm d}^3\vec{q}}{(2\pi)^3}
 \frac{e^{i q_3 r} + e^{-iq_3 r} - 2}{2}
 \biggl\{ 
 \frac{\tau^2 T}{\vec{q}^2 + \Pi_E(0,\vec{q})}
 + 
 T \sum_{\qqbo\neq 0}
 \Bigl(
   2 - e^{i \qqbo\tau} - e^{- i \qqbo \tau} 
 \Bigr)
 \times \nn 
 \!\! & \times & \!\! 
 \biggl[ 
 \biggl( 
   \frac{1}{\qqbo^2} + \frac{1}{\vec{q}^2}
 \biggr)  
 \frac{1}{\qqbo^2 + \vec{q}^2 + \Pi_E(\qqbo,\vec{q}) }
 + 
 \biggl( 
  \frac{1}{q_3^2} - \frac{1}{\vec{q}^2}
 \biggr) 
 \frac{1}{\qqbo^2 + \vec{q}^2 + \Pi_T(\qqbo,\vec{q}) }
 \biggr] 
 \biggr\}
 \;. \la{eqA}
\ea
We have kept the spatial exponents 
in a form which makes it clear that the 
divergent-looking term $1/q_3^2$ is in fact harmless.

In order to carry out the remaining sum in \eq\nr{eqA}, it is useful
to write the resummed propagators in a spectral representation. For any 
function $\Delta(i\qqbo)$ which behaves regularly enough at infinity, 
we can define the spectral density $\rho(q^0)$ through
\be
 \rho(q^0) \equiv \frac{1}{2 i} 
 \Bigl[ 
   \Delta(i\qqbo \to q^0 + i 0^+) - \Delta(i\qqbo \to q^0 - i 0^+)
 \Bigr]
 \;, \la{rhodef}
\ee
which leads to the inverse transform
\be
 \Delta(i\qqbo) = \int_{-\infty}^{\infty} \! \frac{{\rm d} q^0}{\pi}
 \frac{\rho(q^0)}{q^0 - i \qqbo}
 \;.
\ee
In particular, 
\be
 \frac{1}{\qqbo^2 + \vec{q}^2 + \Pi_{E}(\qqbo,\vec{q})}
 = 
 \int_{-\infty}^{\infty} \! \frac{{\rm d} q^0}{\pi}
 \frac{\rho_{E}(q^0)}{q^0 - i \qqbo}
 \;,
 \la{spectral}
\ee
and similarly with $\Pi_T$. 
The properties of the spectral functions $\rho_T, \rho_E$ are 
described in \eqs\nr{PropR}--\nr{rhoE} of Appendix~B.

Inserting the spectral representations into \eq\nr{eqA},
and making use of \eq\nr{bsum},  
we can carry out the remaining sums: 
for $0 < \tau < \beta$, 
\ba
 T \sum_{\qqbo\neq 0} 
 \frac{2 - e^{i \qqbo \tau} - e^{-i \qqbo \tau}}{\qqbo^2 ({q^0 - i \qqbo})} 
  & = &
  \frac{8 T}{q^0}
  \sum_{n=1}^{\infty} \sin^2 ( \pi n T ) 
  \biggl[
   \frac{1}{(2\pi n T)^2 } - 
   \frac{1}{(2\pi n T)^2  + (q^0)^2}
  \biggr]
  \nn 
  & = & 
 \frac{\tau (\beta - \tau)}{q^0 \beta} - 
 \frac{\nB(q^0)}{(q^0)^2}
 \Bigl[ 
  1 + e^{\beta q^0} - e^{\tau q^0}- e^{(\beta - \tau) q^0}
 \Bigr]
 \;, \la{sum1} \\
  T \sum_{\qqbo\neq 0} 
  \frac{2 - e^{i \qqbo \tau} - e^{-i \qqbo \tau}}{q^0 - i \qqbo} 
  & = & 
  8 T q^0 
  \sum_{n=1}^{\infty} \sin^2 ( \pi n T ) 
   \frac{1}{(2\pi n T)^2  + (q^0)^2}
  \nn 
  & = & 
 {\nB(q^0)}
 \Bigl[ 
  1 + e^{\beta q^0} - e^{\tau q^0}- e^{(\beta - \tau) q^0}
 \Bigr]
 \;, \la{sum2}
\ea
where $\nB(q^0)\equiv 1/[\exp(\beta q^0) - 1]$. Noting furthermore 
that (cf.\ \eq\nr{spectral})
\be
 \int_{-\infty}^{\infty} \! \frac{{\rm d} q^0}{\pi}
 \frac{\rho_{E}(q^0)}{q^0}
  =
 \frac{1}{\vec{q}^2 + \Pi_{E}(0,\vec{q})}
 \;, \la{sumrule}
\ee
we arrive at 
\ba
 C_E^{(2)}(\tau,r) 
 & = & g^2 C_F \int \! \frac{{\rm d}^3\vec{q}}{(2\pi)^3}
 \frac{e^{i q_3 r} + e^{-iq_3 r} - 2}{2}
 \biggl\{ 
 \frac{\tau}{\vec{q}^2 + \Pi_E(0,\vec{q})}
 + 
 \nn & + & 
 \int_{-\infty}^{\infty} \! \frac{{\rm d} q^0}{\pi}
 {\nB(q^0)}
 \Bigl[ 
  1 + e^{\beta q^0} - e^{\tau q^0}- e^{(\beta - \tau) q^0}
 \Bigr]
  \times \nn 
 & \times & 
 \biggl[ 
 \biggl( 
   \frac{1}{\vec{q}^2} - \frac{1}{(q^0)^2} 
 \biggr)  
 \rho_E(q^0,\vec{q}) 
 + 
 \biggl( 
  \frac{1}{q_3^2} - \frac{1}{\vec{q}^2}
 \biggr) 
 \rho_T(q^0,\vec{q}) 
 \biggr] 
 \biggr\}
 \;. \la{eqB}
\ea
It is useful to note that both the second and the third
row in \eq\nr{eqB} are odd in $q^0$, so that the product is even, 
and the integral could also be restricted to positive values of $q^0$.

\subsection{Equation of motion in Minkowski time}

\eq\nr{eqB} has a form which can directly be
analytically continued to Minkowski time, $\tau\to it$.
Thereby we arrive at the object
we are interested in, $C_{>}(t,r)$ (cf.\ \eqs\nr{bL}, \nr{bE}):
\ba
 C^{(0)}_{>}(t,r) & = & 
 1
 \;, \la{eqE0} \\ 
 C^{(2)}_{>}(t,r) & = & 
 g^2 C_F 
 \int \! \frac{{\rm d}^3\vec{q}}{(2\pi)^3}
 \frac{e^{i q_3 r} + e^{-iq_3 r} - 2}{2}
 \biggl\{ 
 \frac{i t}{\vec{q}^2 + \Pi_E(0,\vec{q})} 
 + 
 \nn & + & 
 \int_{-\infty}^{\infty} \! \frac{{\rm d} q^0}{\pi}
 {\nB(q^0)}
 \biggl[ 
   1 +  e^{\beta q^0} 
   - e^{i q^0 t}  - e^{\beta q^0} e^{-i q^0 t}
 \biggr] 
  \times \nn 
 & \times & 
 \biggl[ 
 \biggl( 
   \frac{1}{\vec{q}^2} - \frac{1}{(q^0)^2} 
 \biggr)  
 \rho_E(q^0,\vec{q}) 
 + 
 \biggl( 
  \frac{1}{q_3^2} - \frac{1}{\vec{q}^2}
 \biggr) 
 \rho_T(q^0,\vec{q}) 
 \biggr] 
 \biggr\}
 \;. \la{eqE}
\ea

We are now in a position to identify the heavy-quark potential. 
We define it as the quantity which determines the time behaviour of our 
correlator, in accordance with the Schr\"odinger equation: 
\be
 i \partial_t C_{>}(t,r) \equiv V_{>}(t,r) C_{>}(t,r)
 \;.  \la{Seq2}
\ee
Note that this $V_{>}(t,r)$ depends, in general, on the time $t$ and 
on the time-ordering that is used for defining 
the correlator $C_{>}(t,r)$.

In perturbation theory, \eq\nr{Seq2} needs to be fulfilled order 
by order. This implies that 
\ba
  i \partial_t C_{>}^{(0)}(t,r) & = & 0
  \;, \\
  i \partial_t C_{>}^{(2)}(t,r) & = & V_{>}^{(2)}(t,r) C_{>}^{(0)}(t,r)
  \;. \la{Spert}
\ea
Eqs.~\nr{eqE0}, \nr{eqE} then immediately lead to the identification 
\ba
 V_{>}^{(2)}(t,r) & = & 
 g^2 C_F
 \int \! \frac{{\rm d}^3\vec{q}}{(2\pi)^3}
 \frac{2 - e^{i q_3 r} - e^{-iq_3 r}}{2}
 \biggl\{ 
 \frac{1}{\vec{q}^2 + \Pi_E(0,\vec{q})} 
 + 
 \nn & + & 
 \int_{-\infty}^{\infty} \! \frac{{\rm d} q^0}{\pi}
 {\nB(q^0)} q^0 
 \Bigl[ 
  e^{\beta q^0} e^{-i q^0 t}
   - e^{i q^0 t}  
 \Bigr] 
  \times \nn 
 & \times & 
 \biggl[ 
 \biggl( 
   \frac{1}{\vec{q}^2} - \frac{1}{(q^0)^2} 
 \biggr)  
 \rho_E(q^0,\vec{q}) 
 + 
 \biggl( 
  \frac{1}{q_3^2} - \frac{1}{\vec{q}^2}
 \biggr) 
 \rho_T(q^0,\vec{q}) 
 \biggr] 
 \biggr\}
 \;. \la{Vtr}
\ea
For $V_{<}^{(2)}(t,r)$, the analogue of \eq\nr{Seq2}
reads (cf.\ \eq\nr{Seqprime})
\be
 i \partial_t C_{<}(t,r) \equiv - V_{<}(t,r) C_{<}(t,r)
 \;,  \la{Seq3}
\ee
where the symmetry mentioned below \eq\nr{simplelarger}, or 
an explicit computation with NRQCD, 
yield $V_{<}^{(2)}(t,r) = V_{>}^{(2)}(-t,r)$.

%
\section{Discussion of the main results}
\la{se:discussion}

We have shown in the previous section that 
the real-time ``Wilson loop'' $C_{>}(t,r)$, characterising the propagation of 
two infinitely heavy quarks separated by a distance $r$ at 
a finite temperature $T$, satisfies \eq\nr{Seq2}, where 
the potential $V_{>}(t,r)$ is given  in \eq\nr{Vtr} 
to first non-trivial order in $g^2$. 
Let us now discuss the behaviour of $V_{>}(t,r)$
in various limits. 

Perhaps the most interesting limit is what happens 
at large times, ${t\to\infty}$. In this limit, the exponential
functions $\exp(\pm i q^0 t)$ average to zero, unless they are
multiplied by a singular prefactor, $\sim 1/q^0$: then 
\be
 \lim_{t\to\infty} \frac{e^{i q^0 t}-e^{-i q^0 t}}{q^0} 
 = 2 \pi i \; \delta(q^0)
 \;. 
\ee
Given that $\nB(q^0) \approx 1/\beta q^0$ for $|q^0| \ll T$, 
and that $\rho_E(q^0,\vec{q})$ has a term linear in $q^0$, cf.\ 
\eq\nr{rhoE}, such a $\delta$-function indeed emerges from the 
part containing $\rho_E(q^0,\vec{q})$. We obtain
\ba
 \lim_{t\to \infty} V_{>}^{(2)}(t,r) & = & 
 g^2 C_F
 \int \! \frac{{\rm d}^3\vec{q}}{(2\pi)^3}
 \Bigl( 1 - e^{i q_3 r} \Bigr)  
 \biggl\{ 
 \frac{1}{\vec{q}^2 + \mD^2} 
 - \frac{i \pi \mD^2}{\beta}
 \frac{1}{|\vec{q}|(\vec{q}^2 + \mD^2)^2} \biggr\}
 \la{tinfty} \\[3mm] 
 & = & 
 -\frac{g^2 C_F}{4\pi} \biggl[ 
 \mD + \frac{\exp(-\mD r)}{r}
 \biggr] - \frac{i g^2 T C_F}{4\pi} \, \phi(\mD r)	
 \;, \la{expl}
\ea
where we have also made use of \eq\nr{smallPiE}, 
and $\mD$ is the Debye mass parameter. The $r$-independent constant 
in \eq\nr{expl} has been evaluated in dimensional regularization, 
while the dimensionless function $\phi(x)$, 
\be
 \phi(x) = 
 2 \int_0^\infty \! \frac{{\rm d} z \, z}{(z^2 +1)^2}
 \biggl[
   1 - \frac{\sin(z x)}{zx} 
 \biggr]
 \;,
\ee
is finite and strictly increasing, 
with the limiting values $\phi(0) = 0$, $\phi(\infty) = 1$.\footnote{%
 The latter part of the 
 integral can be expressed as Meijer's $G$-function,
 with certain arguments.} 

We observe, thus, that apart from a standard Debye-screened
static potential, the potential $V_{>}(t,r)$ also has an imaginary part.
The sign is such that \eq\nr{Seq2} leads to exponential decay
for $C_{>}(t,r)$. 
The imaginary part obtained for $V_{<}(t,r)$ has the same 
magnitude but opposite sign, so  
that \eq\nr{Seq3} leads to exponential decay as well.
We will refer to the imaginary part as the ``decay width''.

In order to get a rough impression of the numerical 
orders of magnitude, let us solve the time-independent
Schr\"odinger equation with the real part of \eq\nr{tinfty}:
\be
 \biggl[
  -\frac{1}{M} 
 \biggl(
   \frac{{\rm d}^2}{{\rm d} r^2} + \frac{2}{r} 
   \frac{{\rm d}}{{\rm d} r}
 \biggr) 
 - g^2 C_F \frac{\exp(-\mD r)}{4\pi r}
 \biggr] \psi(r) 
 = 
 E\, \psi(r) 
 \;, \la{simpS}
\ee 
where $E$ denotes the binding energy.
If $M$ is very large, $g^2 C_F M/8 \pi \gg \mD$, Debye screening 
can be ignored, and we get the regular hydrogen atom solution, 
with an inverse Bohr radius $1/r_0 = g^2 C_F M/8 \pi$, and a binding 
energy $|E_0| = M (g^2 C_F /8 \pi)^2$.  
Treating the imaginary part, $\Gamma$, as a perturbation, 
$\Gamma \simeq  (g^2 T C_F / 4\pi) \int\!{\rm d}^3 \vec{r} \, 
|\psi(r)|^2 \, \phi(\mD r)$, 
it is clear that $|\Gamma| \le  g^2 T C_F / 4\pi$,
because $0\le \phi(\mD r) \le 1$; 
in fact, for $r_0 \mD \ll 1$ as was the assumption,  
$|\Gamma| \ll  g^2 T C_F / 4\pi$. We thus note that
for a very heavy quark mass $M$, the decay width can equal 
the binding energy only at temperatures $T \gg g^2 C_F M / 16 \pi$.

Lowering towards more realistic quark masses at a fixed temperature, 
$r_0$ increases, and at some point the Debye radius $1/\mD$ becomes 
important. Consequently the binding energy decreases fast and, 
simultaneously, the width $\Gamma$
increases, because the wave function spreads out, 
whereby $\phi(\mD r)$ obtains non-zero values. 

Unfortunately, it is non-trivial to make these arguments precise: 
to fix the scale parameter appearing in $g^2$ and $\mD^2$, 
another order in the perturbative
expansion would be needed; once the width is substantial, it can no 
longer be treated as a perturbation; and the full time-dependence 
of $V_{>}^{(2)}(t,r)$ should be considered in order 
to determine the width reliably. 
Nevertheless, for illustration, we may fix the scales inside
$g^2$ and $\mD^2$ by making use of another context where several
orders are available, namely that of dimensionally reduced
effective theories~\cite{dr,generic,bn}. Concretely, employing
simple analytic expressions that can be extracted from Ref.~\cite{adjoint}, 
\be
 g^2 \simeq \frac{8\pi^2}{9 \ln( 9.082\, T/ \Lambdamsbar)} 
 \;, \quad
 \mD^2 \simeq \frac{4\pi^2 T^2}{3 \ln(7.547\, T/\Lambdamsbar)}
 \;,
 \qquad \mbox{for $\Nc = \Nf = 3$}
 \;; \la{numg2}
\ee 
solving numerically the Schr\"odinger-equation in 
\eq\nr{simpS}; and treating the imaginary part still as 
a perturbation, we obtain \fig\ref{fig:b} for the bottomonium
system. (For charmonium the relevant mass 
$M\simeq 1.25$~GeV is quite a bit smaller, and moves the 
patterns to smaller temperatures, where our methods 
are no longer reliable.)
We stress that this figure is not meant to be  
quantitatively accurate, but it does illustrate something about
the qualitative behaviour of the system: namely, that the 
width increases with temperature; the binding energy decreases; 
and the width exceeds the binding energy already much before
the latter disappears (i.e.\ bottomonium ``melts''). 

\begin{figure}[t]


\centerline{%
\epsfysize=7.0cm\epsfbox{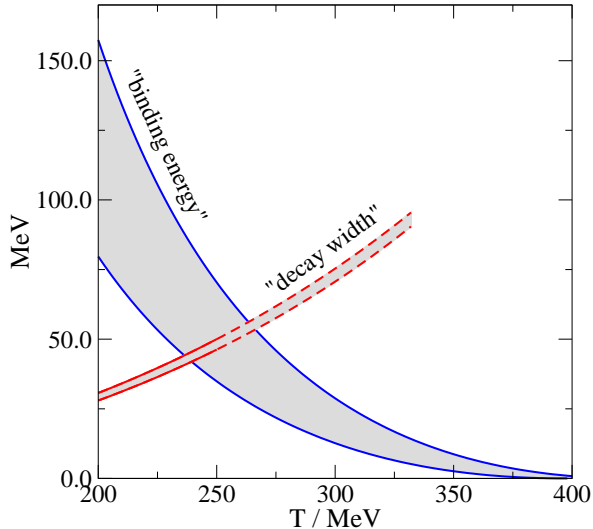}%
}

\caption[a]{\small Very rough estimates for the binding energy
and decay width of the bottomonium system ($M \simeq 4.25$~GeV), 
as a function of the temperature, 
based on \eqs\nr{tinfty}--\nr{numg2}.
The error bands have been obtained by varying $\Lambda_\tinymsbar^{(3)}$
from 300~MeV (lower edges) to 450~MeV (upper edges). We expect 
the true errors to be much larger than these bands,  
for reasons discussed in the text.
} 
\la{fig:b}
\end{figure}

Let us at this point inspect how the results would 
change in the {\em classical limit}.\footnote{
 This limit is of course singular in many respects, 
 but can still be formally defined~\cite{hsqr}, 
 and allows to address numerically many important
 non-perturbative problems in weakly-coupled
 gauge theories~\cite{oldclas}--\cite{instab}, 
 in spite of the presence of significant 
 discretization artifacts~\cite{cllattice,bl} 
 in the classical lattice gauge theory
 simulations~\cite{oldclas} that are employed in the thermal context.
 }
For this, we need to reintroduce $\hbar$. 
After the standard rescaling of the fields, $\hbar$ appears in connection 
with the coupling constant, as $g^2\hbar$, as well as in the extent of 
the Euclidean time direction, as $\beta \hbar$. In the classical limit, 
therefore, $\nB(q^0) \to 1/\beta\hbar q^0$.

We now note that $\hbar$ cancels from the 
imaginary part of \eq\nr{expl}, where the combination
$g^2\hbar / \beta \hbar = g^2T$ appears. This means that 
the decay width exists also in the classical limit. In fact, 
nothing but the decay width exists in the classical limit!
Moreover, time-ordering has no meaning in the classical limit, 
which explains why the imaginary parts have the same 
magnitude in $V_{>}^{(2)}(t,r)$ and $V_{<}^{(2)}(t,r)$.
That a decay width / damping rate should be classical, 
need not be surprising: particularly detailed demonstrations 
of this have been provided for scalar field theory~\cite{sft}. 

Finally we remark that for $T\to 0$, i.e. $\beta\to\infty$,
we can restrict to $q^0 > 0$ in \eq\nr{Vtr} because of the symmetry, and 
then  $\nB(q^0)\to 0$, 
$\nB(q^0)\exp(\beta q^0)\to 1$, whereby the structure of \eq\nr{Vtr}
simplifies. In particular, there is no imaginary part 
any more for $t\to\infty$, only the standard Coulomb-potential.

%
\section{Conclusions and Outlook}
\la{se:concl}

The purpose of this note has been to present a derivation
of a ``static potential'' that a heavy quark-antiquark pair
propagating in a thermal medium feels. An integral representation 
of the result is shown in \eq\nr{Vtr}. This potential could
then be inserted into \eq\nr{Seq}, in order to estimate a certain
real-time Green's function for the vector current, whose 
Fourier-transform determines the quarkonium contribution 
to the dilepton production rate.
A detailed numerical evaluation 
is postponed to future work.

It is the first term in \eq\nr{Vtr} which
represents the standard time-independent Debye-screened 
potential. We note that this potential behaves as 
$- g^2 C_F \exp(-\mD r)/4\pi r$ as a function of $r$. Thus it 
differs from the potential extracted from the correlator of 
two Polyakov loops in Euclidean spacetime, which behaves as
$\sim - g^4 \exp(-2 \mD r)/ 4\pi r$ at phenomenologically 
relevant temperatures~\cite{ah}, and as 
$\sim - g^{4+n} \exp( - \mG r)/4\pi r$, $n > 0$,
at asymptotically high temperatures~\cite{MDNLO}, where $\mG$ is the lightest
glueball mass of three-dimensional pure Yang-Mills theory~\cite{mt}, 
with the gauge coupling $g_3^2 = g^2 T[1 + \rmO(g)]$~\cite{gE2,pg2}.

Motivated by our perturbative analysis, one can however envisage
ways of defining a modified (real part of the) static potential, 
which would be gauge-invariant, possess the correct perturbative 
limit, and allow for a direct non-perturbative measurement with
lattice simulations. Inspecting \eq\nr{eqB}, the correct term 
is seen to sit on the first row. The corresponding functional
dependence on $\tau$ differs from that on the second row 
in a qualitative way: for instance, it is linear in $\tau$ 
at $\tau\ll\beta$, while the second row is quadratic; and 
it is non-periodic in $\tau\to \beta-\tau$, while the second 
row is periodic. Both of these functional features can in principle 
also be isolated from non-perturbative data for the Euclidean Wilson 
loop; whether they lead to useful practical recipes for determining 
the real part of the static potential beyond perturbation theory, 
remains however to be tested. 

Apart from the standard potential, we note that \eq\nr{Vtr}
also has another term, the second one,
with a fairly rich structure. In the limit
$t\to \infty$, the second terms amounts to a thermal decay width, 
induced by Landau damping of the low-frequency gauge fields that 
mediate interactions between the two heavy quarks. 
The thermal width induces a certain width also to the quarkonium 
peak in the dilepton production rate, thus making the peak much wider
than at zero temperature. Physically, the Landau damping underlying 
this phenomenon originates from an energy transfer from low-frequency 
gauge fields that are responsible for the static interaction, 
to the ``hard'' particles (which have momenta of the order of 
the temperature) existing in the thermal plasma; technically, 
it originates from the ``cut contribution'' to the gluon spectral 
function at $|q^0| < |{\bf q}|$, cf.\ \eqs\nr{rhoT}, \nr{rhoE}. 
Restricting to a very rough estimate, we have shown that 
this part leads to physically sensible qualitative structures,
namely a width which increases with the temperature, exceeding 
the binding energy already quite a bit before the latter disappears and 
quarkonium ``melts''.

It is worth stressing, however, that apart from the
limiting value at $t\to\infty$, the potential has a non-trivial 
time dependence, which also plays a role 
in the solution of \eq\nr{Seq} and, consequently, for 
the quarkonium contribution to the dilepton production rate. 

In order to define the significance of these findings for 
heavy quarkonium, its finite mass $M$ has to 
be taken into account systematically. 
For this one can make use of NRQCD~\cite{nrqcd}, 
allowing to account analytically for all exponential effects 
like  $\exp(-2 \beta M )$, so that one can concentrate 
on the softer dynamics around the threshold. Perturbatively, 
this amounts to the solution of \eq\nr{Seq} with $V_{>}^{(2)}(t,r)$
inserted inside the round brackets. 
However, one might also be able to go beyond perturbation theory, 
by making use of the classical approximation for the gauge 
fields appearing in NRQCD, provided that genuinely quantum effects, 
such as the first term in \eq\nr{Vtr}, are properly taken into account. 
Hopefully this recipe could be formalised such that one would
be able to account for the orders $(g^2\hbar)^0$ and 
$(g^2\hbar)^1$ in the ``zero-temperature'' expansion, 
but in each case for all orders in the possibly larger
finite-temperature expansion parameters, $g^2 T/\mD$ and $g^2 T/\mG$.

\vspace*{-0.5cm}

%
\section*{Acknowledgements}

This work was partially supported by the BMBF project
{\em Hot Nuclear Matter from Heavy Ion Collisions 
     and its Understanding from QCD}.
P.R.\ was partially supported by the US Department of Energy, 
grant number DE-FG02-00ER41132.



\appendix
\renewcommand{\thesection}{Appendix~\Alph{section}}
\renewcommand{\thesubsection}{\Alph{section}.\arabic{subsection}}
\renewcommand{\theequation}{\Alph{section}.\arabic{equation}}

%
\section{Real-time Green's functions}

\newcommand{\A}{\hat\phi^\mu}
\renewcommand{\B}{\hat{\phi}^\nu}
\newcommand{\I}{\int\!{\rm d}t\,{\rm d}^3 \vec{x}\,e^{i Q\cdot x}}

For completeness and to fix the notation, 
we list in this Appendix some common definitions and relations 
that apply to two-point correlation functions built out of bosonic 
operators; for more details see, e.g., Refs.~\cite{leb,kg}. 

We use the notation introduced at the beginning of 
\se\ref{se:details}, with $t\equiv x^0$ and $\tau \equiv \tilde x_0$.
Arguments of operators
denote implicitly whether we are in Minkowskian or Euclidean space-time.
In particular, Heisenberg-operators are defined as
\be
 \hat O(t,\vec{x}) \equiv 
 e^{i \hat H t} \hat O(0,\vec{x}) e^{- i \hat H t}
 \;, \quad
 \hat O(\tau,\vec{x}) \equiv 
 e^{\hat H \tau} \hat O(0,\vec{x}) e^{- \hat H \tau}
 \;.
\ee
The thermal ensemble is defined by the density matrix
$\hat\rho = \mathcal{Z}^{-1} \exp(-\beta \hat H)$.
We denote the operators which appear in the two-point functions
by $\A(x)$, $\B(x)$; they could
either be elementary field operators or composite operators.

We can now define various classes of 
correlation functions. The ``physical'' correlators are defined as
\ba
 \tilde C_{>}^{\mu\nu}(Q) & \equiv & 
 \I \Bigl\langle \A(x) \B(0) \Bigl\rangle
 \;,   
 \la{bL}
 \\
 \tilde C_{<}^{\mu\nu}(Q) & \equiv & 
 \I \Bigl\langle \B(0)  \A(x) \Bigl\rangle
 \;,   
 \la{bS}
 \\
 \rho^{\mu\nu}(Q) & \equiv & 
 \I \Bigl\langle \fr12 \Bigl[ \A(x) , \B(0) \Bigr] \Bigl\rangle
 \;,   
 \la{brho} \\
 \tilde \Delta^{\mu\nu}(Q) & \equiv & 
 \I \Bigl\langle \fr12 \Bigl\{ \A(x) , \B(0) \Bigr\} \Bigl\rangle
 \;,   
 \la{bdelta}
\ea
where $\rho^{\mu\nu}$ is called the spectral function, 
while the ``retarded''/``advanced'' correlators can be defined as 
\ba
 \tilde C_R^{\mu\nu}(Q) & \equiv & 
 i \I \Bigl\langle \Bigl[ \A(x) , \B(0) \Bigr] \theta(t) \Bigl\rangle
 \;, 
 \la{bR}
 \\
 \tilde C_A^{\mu\nu}(Q) & \equiv & 
 i \I \Bigl\langle  - \Bigl[ \A(x) , \B(0) \Bigr] \theta(-t) \Bigl\rangle
 \;. 
 \la{bA} 
\ea
On the other hand, from the computational point of view one 
is often faced with ``time-ordered'' correlation functions, 
\ba
 \tilde C_T^{\mu\nu}(Q) & \equiv & 
  \I \Bigl\langle \A(x)  \B(0) \theta(t)
                 +\B(0)  \A(x) \theta(-t) \Bigl\rangle
 \;, \la{bT}
\ea
which appear in time-dependent perturbation theory, or with 
the ``Euclidean'' correlator
\ba
 \tilde C_E^{\mu\nu}(\tilde Q) & \equiv & 
 \int_0^\beta\!{\rm d}\tau \int \! {\rm d}^3 \vec{x}\,
  e^{i \tilde Q \cdot \tilde x}
  \Bigl\langle \A(\tilde x)  \B(0) \Bigl\rangle
 \;,
 \la{bE}
\ea
which appears in non-perturbative formulations.
Note that the Euclidean correlator is also time-ordered by definition, 
and can be computed with Euclidean functional integrals. 

Now, all of the correlation functions defined 
can be related to each other.
In particular, all correlators can be expressed in terms of the spectral
function, which in turn can be determined as a certain analytic 
continuation of the Euclidean correlator. In order to do this, 
we may first insert sets of energy eigenstates, to obtain
the Fourier-space version of the so-called Kubo-Martin-Schwinger (KMS) 
relation:  
$
 \tilde C_{<}^{\mu\nu}(Q) 
 = e^{-\beta q^0} \tilde C_{>}^{\mu\nu}(Q)
$.
Then 
$
 \rho^{\mu\nu}(Q)
 = [\tilde C_{>}^{\mu\nu}(Q) - \tilde C_{<}^{\mu\nu}(Q)]/2
$ and, conversely, 
\be
 \tilde C_{>}^{\mu\nu}(Q) = 2[1 + \nB(q^0)] \rho^{\mu\nu}(Q)
 \;, \quad
 \tilde C_{<}^{\mu\nu}(Q) = 2 \nB(q^0) \rho^{\mu\nu}(Q)
 \;. \la{bLSrel}
\ee
Moreover, 
$
 \tilde \Delta^{\mu\nu}(Q) = [1 + 2 \nB(q^0)] \rho^{\mu\nu}(Q)
$.
Inserting the representation
\be
 \theta(t) = i \int_{-\infty}^{\infty} \! \frac{{\rm d}\omega}{2\pi}
 \frac{e^{-i\omega t}}{\omega + i 0^+}
 \la{theta}
\ee
into the definitions of $\tilde C_R$, $\tilde C_A$, we obtain
\be
 \tilde C_R^{\mu\nu}(Q) 
 =  \int_{-\infty}^{\infty} \! \frac{{\rm d}\omega}{\pi} 
 \frac{\rho^{\mu\nu}(\omega,\vec{q})}{\omega -q^0- i 0^+}
 \;, \quad 
 \tilde C_A^{\mu\nu}(Q)
 =  \int_{-\infty}^{\infty} \! \frac{{\rm d}\omega}{\pi} 
 \frac{\rho^{\mu\nu}(\omega,\vec{q})}{\omega -q^0+ i 0^+}
 \;. 
\ee
Doing the same with $\tilde C_T$ and making use of 
\be
 \frac{1}{\Delta \pm i0^+} = P\Bigl(\frac{1}{\Delta}\Bigr) 
 \mp i \pi \delta(\Delta)
 \;,  \la{delta}
\ee
produces
\be
 \tilde C_T^{\mu\nu}(Q) = 
 \int_{-\infty}^{\infty} \! \frac{{\rm d}\omega}{\pi} 
 \frac{i\rho^{\mu\nu}(\omega,\vec{q})}{q^0 - \omega + i 0^+} + 
 2 \rho^{\mu\nu}(q^0,\vec{q}) \nB(q^0)
 \;. 
\ee
Finally, writing the argument inside the $\tau$-integration
in \eq\nr{bE} as a Wick rotation of the integrand in \eq\nr{bL}, 
which in turn is expressed as an inverse Fourier transform 
of $\tilde C_{>}^{\mu\nu}(Q)$,
for which \eq\nr{bLSrel} is inserted, 
and changing orders of integration, we get 
\ba
 \tilde C_E^{\mu\nu}(\tilde Q) & = & 
 \int_0^\beta \! {\rm d}\tau\, e^{i \qq \tau}
 \int_{-\infty}^{\infty} \frac{{\rm d}\omega }{2\pi} e^{-\omega \tau} 
 \tilde C_{>}^{\mu\nu}(\omega,\vec{q}) =
 \int_{-\infty}^{\infty} \! \frac{{\rm d}\omega}{\pi} 
 \frac{\rho^{\mu\nu}(\omega,\vec{q})}{\omega - i \qq}
 \;. \la{bErhorel}
\ea
This relation can formally be 
inverted by making use of \eq\nr{delta}, 
\be
 \rho^{\mu\nu}(q^0,\vec{q}) 
 = \frac{1}{2i} \Bigl[ 
 \tilde C_E^{\mu\nu}(-i[q^0+i 0^+],\vec{q}) - 
 \tilde C_E^{\mu\nu}(-i[q^0-i 0^+],\vec{q})
 \Bigr]
 \;. \la{Discdef}
\ee

We also recall that bosonic Matsubara sums can be carried out through
\ba
  T \sum_{\qqbo} 
 \frac{i \qqbo c + d}
 {\qqbo^2 + E^2} e^{i \qqbo \tau}
 & \equiv &  
 (c\partial_\tau + d) 
  T \sum_{\qqbo} 
 \frac{e^{i \qqbo \tau}}
 {\qqbo^2 + E^2} 
 \\ 
 & = & 
 \frac{\nB(E)}{2 E} \Bigl[ 
  (-cE + d) e^{(\beta - \tau) E} + (cE + d) e ^{\tau E}
 \Bigr] \;,
 \la{bsum}
\ea
where $\qqbo = 2 \pi n T$,
with $n$ an integer, and we assumed $0 <  \tau < \beta$. 
This equation can be used as a starting point for determining the sums in 
\eqs\nr{sum1}, \nr{sum2}.

%
\section{Resummed gluon propagator}

Introducing the projection operators 
(see, e.g., Refs.~\cite{leb,kg})
\ba
 P^T_{00}(\tilde Q) \!\! & = & \!\! 
 P^T_{0i}(\tilde Q) = P^T_{i0}(\tilde Q) \equiv 0
 \;, \quad P^T_{ij}(\tilde Q) \equiv \delta_{ij} - 
     \frac{\tilde q_i \tilde q_j}{\tilde\vec{q}^2}
 \;, \la{PT} \\ 
 P^E_{\mu\nu}(\tilde Q) \!\! & \equiv & \!\!
 \delta_{\mu\nu} - \frac{\tilde q_\mu \tilde q_\nu}{\tilde Q^2}
 - P^T_{\mu\nu}(\tilde Q)
 \;, \la{PE}
\ea
the Euclidean gluon propagator can be written as 
\be
 \langle A^a_{\mu} (\tilde x) A^b_\nu (\tilde y) \rangle = 
 \delta^{ab} \Tint{\tilde Q} e^{i\tilde Q\cdot (\tilde x - \tilde y)}
 \biggl[
   \frac{P^T_{\mu\nu}(\tilde Q)}{\tilde Q^2 + \Pi_T(\tilde Q)} + 
   \frac{P^E_{\mu\nu}(\tilde Q)}{\tilde Q^2 + \Pi_E(\tilde Q)} + 
   \xi \frac{\tilde q_\mu \tilde q_\nu}{(\tilde Q^2)^2} 
 \biggr] 
 \;,  \la{prop}
\ee
where $\xi$ is the gauge parameter. The Hard Thermal 
Loop~\cite{htlold,htl} contributions read
\ba
 \Pi_T(\tilde Q) & = & 
 \frac{\mD^2}{2} \biggl\{ 
  \frac{(i\qqbo)^2}{\tilde\vec{q}^2} + 
  \frac{i \qqbo}{2 |\tilde\vec{q}|}
  \biggl[
   1 - \frac{(i \qqbo)^2}{\tilde\vec{q}^2} 
  \biggr]
  \ln \frac{i\qqbo + |\tilde\vec{q}|}{i\qqbo - |\tilde\vec{q}|} 
 \biggr\}
  \;, \\
 \Pi_E(\tilde Q) & = & 
 \mD^2 
  \biggl[
   1 - \frac{(i \qqbo)^2}{\tilde\vec{q}^2} 
  \biggr]
  \biggl[ 
    1 -  \frac{i \qqbo}{2 |\tilde\vec{q}|} 
    \ln \frac{i\qqbo + |\tilde\vec{q}|}{i\qqbo - |\tilde\vec{q}|} 
  \biggr]
  \;,
\ea
where $\qqbo$ denotes bosonic Matsubara frequencies, and 
\be
 \mD^2 = g^2 T^2 \biggl( \frac{\Nc}{3} + \frac{\Nf}{6} \biggr) 
 \;.
\ee
In the limit $i\qqbo\to 0$ but with $|\tilde\vec{q}|\neq 0$, 
$\Pi_T \to 0$, $\Pi_E \to \mD^2$, while
for $|\tilde\vec{q}| \to 0$ with $i \qqbo \neq 0$, 
$\Pi_T$, $\Pi_E \to \mD^2/3$.

After analytic continuation, $i\qqbo \to q^0 + i 0^+$, 
the propagators become 
\be
 \frac{1}{\tilde Q^2 + \Pi_{T(E)}(\qqbo,\tilde\vec{q})} \to 
 \frac{1}{-(q^0 + i 0^+)^2 + \vec{q}^2 + 
 \Pi_{T(E)}(-i(q^0 + i 0^+),\vec{q})}
 \;, \la{PropR}
\ee
where
\ba
 \Pi_T(-i(q^0 + i 0^+),\vec{q}) & = & 
 \frac{\mD^2}{2} 
 \biggl\{ 
   \frac{(q^0)^2}{\vec{q}^2} + 
   \frac{q^0}{2|\vec{q}|}
   \biggl[
     1 -  \frac{(q^0)^2}{\vec{q}^2}
   \biggr] 
   \ln\frac{q^0 + i 0^+ + |\vec{q}|}{q^0 + i 0^+ - |\vec{q}|}
 \biggr\} 
 \;, \\
 \Pi_E(-i(q^0 + i 0^+),\vec{q}) & = & 
 \mD^2 
   \biggl[
     1 -  \frac{(q^0)^2}{\vec{q}^2}
   \biggr] 
   \biggl[
     1 -     
     \frac{q^0}{2|\vec{q}|}
   \ln\frac{q^0 + i 0^+ + |\vec{q}|}{q^0 + i 0^+ - |\vec{q}|}
   \biggr] 
 \;.
\ea
For $|q^0| > |\vec{q}|$, $\Pi_T, \Pi_E$ are real. 
For $|q^0| < |\vec{q}|$, they have an imaginary part. 
In particular, for $|q^0| \ll |\vec{q}|$, we get
\ba
 \Pi_T & \approx &  
 \frac{\mD^2}{2} 
 \biggl\{ 
  -i \pi \frac{q^0}{2|\vec{q}|} + 2 \frac{(q^0)^2}{\vec{q}^2} + ...
 \biggr\}
 \;, \la{smallPiT} \\ 
 \Pi_E & \approx &  
 \mD^2
 \biggl\{ 
  1 + i \pi \frac{q^0}{2|\vec{q}|} - 2 \frac{(q^0)^2}{\vec{q}^2} + ...
 \biggr\}
 \;. \la{smallPiE}
\ea
The Hard Thermal Loop resummation of course only describes
the behaviour correctly for $|\vec{q}| \sim gT$; for instance,  
for $|\vec{q}| \sim g^2 T$, loop corrections within the Hard Thermal
Loop effective theory are large, and the small-frequency behaviour 
of the full $\Pi_T$ is determined by a ``colour conductivity'' 
rather than \eq\nr{smallPiT}:  
$\Pi_T \sim - i \sigma q^0$, where $\sigma \sim T/\ln(1/g)$~\cite{dblog,cond}.

The spectral functions $\rho_T, \rho_E$ that are needed in the 
text follow by taking the imaginary part, or discontinuity
(cf.\ \eq\nr{rhodef}), 
of the right-hand side of \eq\nr{PropR}. In particular, 
\ba
 \rho_T(q^0,\vec{q})
 & = & \left\{ 
 \begin{array}{ll} 
 \pi \sign(q^0) \delta 
 \Bigl( 
  (q^0)^2 - \vec{q}^2 - \re\Pi_T
 \Bigr)\;, & |q^0| > |\vec{q}| \\[2mm] \displaystyle 
 \pi \mD^2 \frac{q^0}{4 |\vec{q}|^5} \;, & |q^0| \ll |\vec{q}| 
 \end{array} \right. 
 \;, \la{rhoT} \\[3mm]
 \rho_E(q^0,\vec{q})
 & = & \left\{ 
 \begin{array}{ll} 
 \pi \sign(q^0) \delta 
 \Bigl( 
  (q^0)^2 - \vec{q}^2 - \re\Pi_E
 \Bigr)\;, & |q^0| > |\vec{q}| \\[2mm] \displaystyle 
 - \pi \mD^2 \frac{q^0}{2 |\vec{q}| 
   (\vec{q}^2 + \mD^2)^2} \;, & |q^0| \ll |\vec{q}| 
 \end{array} \right. 
 \;. \la{rhoE} 
\ea
Again, ``soft'' loop corrections to 
these expressions are large for ultrasoft
momenta, $|\vec{q}|\sim g^2 T$. Moreover, loop corrections 
are also significant around the plasmon poles, i.e.\ 
the $\delta$-functions in \eqs\nr{rhoT}, \nr{rhoE}, where 
a finite plasmon decay width gets generated~\cite{bp}.


\end{document}